# IMPACT OF PULSATION ACTIVITY ON THE LIGHT CURVES OF SYMBIOTIC VARIABLES

Marsakova V.I.[1], Andronov I.L.[2], Chinarova L.L.[3], Chyzhyk M.S.[1], Andrych K.D.[1]

[1] Department of Astronomy, Odessa National University
[2] Department "High and Applied Mathematics"
Odessa National Maritime University
[3] Astronomical Observatory, Odessa National University

**Abstract.** We used long-term visual amateur observations of several symbiotic variables for detection of periods that may be caused by pulsation. The examples of multiple periodicities are discussed individually in each case.

Symbiotic variables are long-period interacting binary systems that include evolved red giant and much hotter companion, which, in most systems, is a white dwarf. They show a lot of observational effects associated with both components and their interaction, particularly, mass transfer from a cold to a hot companion. Based on their near-IR characteristics, the symbiotic stars are divided into two main classes [1], depending on whether the colors are stellar-like (S-type) or indicate a thick dust shell (D-type). The majority (≈80%) of cataloged systems are of S-type and have near-IR colors consistent with cool stellar photosphere temperatures of ≈ 3500−4000 K. Most of them have orbital periods of ≈ 500−1000 days. The near-IR colors of the D-type systems indicate the presence of a dust shell which obscures the star and re-emits at longer wavelengths. IR photometric monitoring has shown that these D-type systems have large amplitude variations and that they contain Mira-type variables with pulsation periods in the range of 300–600 days. They are often called the "symbiotic Miras" [2]. Since they must contain the Mira with its dust shell, these D-type systems should have much longer orbital periods than the S-types, a few tens of years and more. The latest review of symbiotic Miras and a comparison with normal Miras can be found in [3]. As mentioned in [4], spectroscopically over a time scale of a few years, the D-type symbiotics have near-infrared velocity behavior typical for Mira variables. In the S-type systems, the variations in the optical photometry are dominated by the mass transfer between the late-type star and the degenerate companion. Extensive time-series of the radial velocities of most S-type symbiotics are dominated by the orbital motion. However, in a few cases, the pulsation of the late-type primary makes a significant contribution to the observed velocities.

Optical and infrared light curves show the very complex behavior that may be described by the such effects [5]:
- high and low activity stages of the hot component,
- flickering caused by accretion disc,
- nova-like outbursts of the hot component (S and D types),
- eclipses,

- reflection effect of red giant (S-type),
- ellipsoidal variability connected with orbital motion (S-type),
- radial pulsations (all D-type and some S-type) and semi-regular variation (S-type) of the cool component,
- long-term dust obscuration (mostly D-type)
- and other types of variability.

Some examples of these variability types are clearly demonstrated by Mikołajewski et al. [6].

We try to investigate the photometric variability of some symbiotic variables using the observations in the visual spectral range. The data for our research was compiled from the international databases AFOEV, VSOLJ and AAVSO and obtained by using the photographic plate collections of the Astronomical Observatory of the Odessa National University and of the Sternberg Astronomical Institute of the Moscow State University.

We have used:
- Periodogram analysis [7, 8];
- Trigonometrical polynomial and multifrequency approximation [9, 10] realized in the program MCV [11];
- "Asymptotic parabolae" [12], "running parabolae" [13] and "running sines" [14];
- Wavelet analysis [15, 16] and scalegram analysis [12].

**CH Cyg**

CH Cyg is well known symbiotic variable [17]. The characteristics of complex optical variability of the classical symbiotic star were analyzed [18]. The long-term ($1840^d$), orbital ($694^d$) periods were obtained. Also small-amplitude ($0.09^m$) pulsations at the period of $99^d$ were confirmed.

**UV Aur**

The carbon variable UV Aur belongs to the S-type, but at the color-color diagram take an intermediate location [19] and is known as a symbiotic Mira. Photometric period of $393\pm0.3^d$ was found in our research that slightly differ from the value listed in catalogue of Belczyński et al. [20]. The phase change and changes in the characteristics of individual cycles for UV Aur were detected as well as the $6800^d$ photometric wave [21]. The hypothesis proposed to explain these changes is that the red giant companion pulsates with a period close (but not equal) to the orbital one. Thus the detected long-period changes in average brightness of UV Aur are explained by proposed model of a binary system with an elliptical orbit and the pulsational -orbital beat. Under these assumptions, the orbital period may be estimated to be $371.5^d$, but this hypothesis has to be checked using spectral observations.

**V1329 Cyg**

In the symbiotic Nova V1329 Cyg, the secondary period of $553^d$ was detected, as well as the $5300^d$ cycle [22]. These waves may be associated with Mira-type pulsations.

**RX Pup**

According to [23], the symbiotic variable RX Pup is composed from a long-period Mira variable surrounded by a thick dust shell and a hot $0{,}8M_\odot$ white dwarf companion. The hot component produces practically all activity observed in the UV, optical and radio range, while variable obscuration of the Mira by circumstellar dust is responsible for long-term changes in the near-infrared magnitudes.

The visual observations (from the AAVSO database) show that RX Pup underwent a nova-like eruption and then it showed continuous weakening of brightness (Fig. 1).

Trying to detect the periodicity, we divided the light curves into three intervals of almost linear changing of the mean brightness. Then we made the periodogram analysis in each interval. The determined periods are listed in the Table 1. As was mentioned by Mikołajewska et al. [23], the pulsational period of 578 days (obtained from the near-IR observations) is not detectable in the visual data. But we have found some other periodicity (of about 350-370, about 285-295 days). One may note, that a half of 578 is 289 days, so a half of the pulsation period is present in all these intervals. We used the observations published in [23] to compare with these visual data (Fig. 2) and could note that several minima and maxima corresponding to the pulsation curve are present in the visual data, but they are significantly narrower and that lead to a half of period. The cyclicity of $350\text{-}370^d$ may be produced by dust clouds around Mira component or accretion structures near hot component.

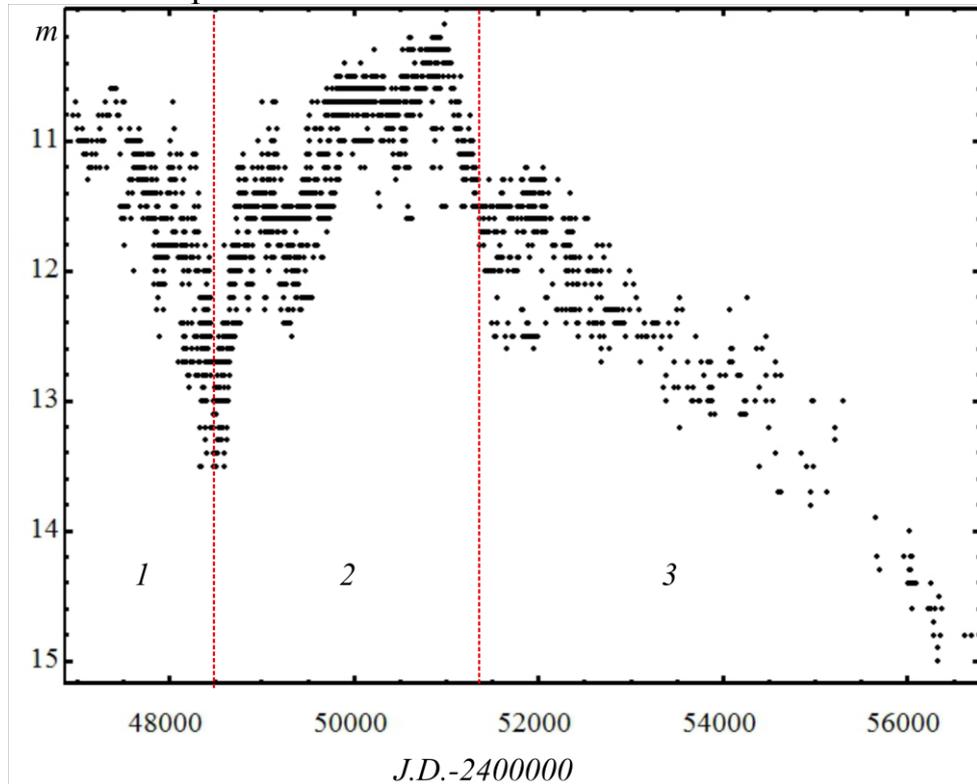

Fig.1. AAVSO's observations of RX Pup.

Table 1. Results of periodogram analysis of AAVSO's observations of RX Pup (P is the period, S is the test function equal to $1-\sigma_{O-C}^2/\sigma_C^2$ [7]).

| Interval 1 | | Interval 2 | | Interval 2 | |
|---|---|---|---|---|---|
| P | S | P | S | P | S |
| 283.7 | 0.09 | 918.7 | 0.26 | 294.3 | 0.09 |
| | | 371.7 | 0.10 | 346.0 | 0.07 |
| | | 286.4 | 0.08 | | |

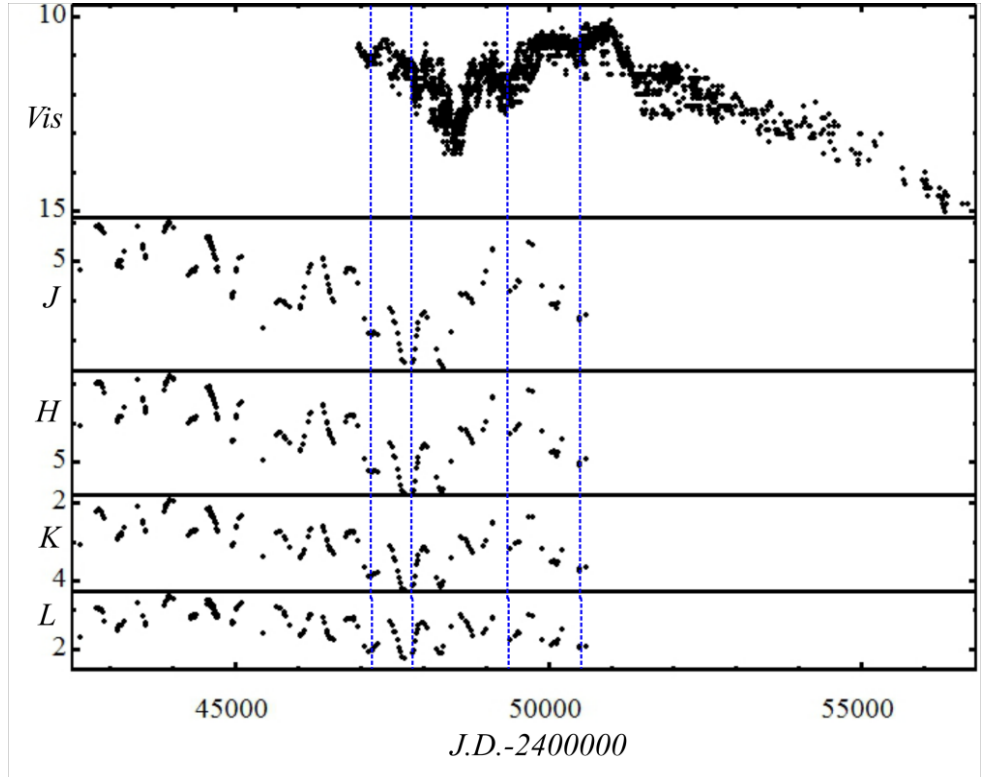

Fig.2. Comparison of visual AAVSO data and infrared observations from [23]. Some minima marked by using blue lines.

Fig. 3-4 illustrate the best multi-frequency approximations of the light curves with different periods and a polynomial trend of second degree that fits long-term changes connected with dust obscurations and nova-like flare.

This example shows that multi-wavelength observations are essential to understand the processes in symbiotic variables.

Table 2. Periods used for approximation of observations of RX Pup in the intervals 1 and 2. ($P$ is the period, $m$ is the degree of trigonometrical polynomial, $S_{O-C}$ is sum of squares of residuals after approximation by using the current period together with the previous ones).

| Interval 1 | | | Interval 2 | | |
|---|---|---|---|---|---|
| P | m | $S_{O-C}$ | P | m | $S_{O-C}$ |
| 578 | 2 | 68,0 | 578 | 2 | 201 |
| 371,7 | 1 | 65,8 | 371,7 | 1 | 180 |
| 918,7 | 1 | 64,5 | 918,7 | 1 | 136 |

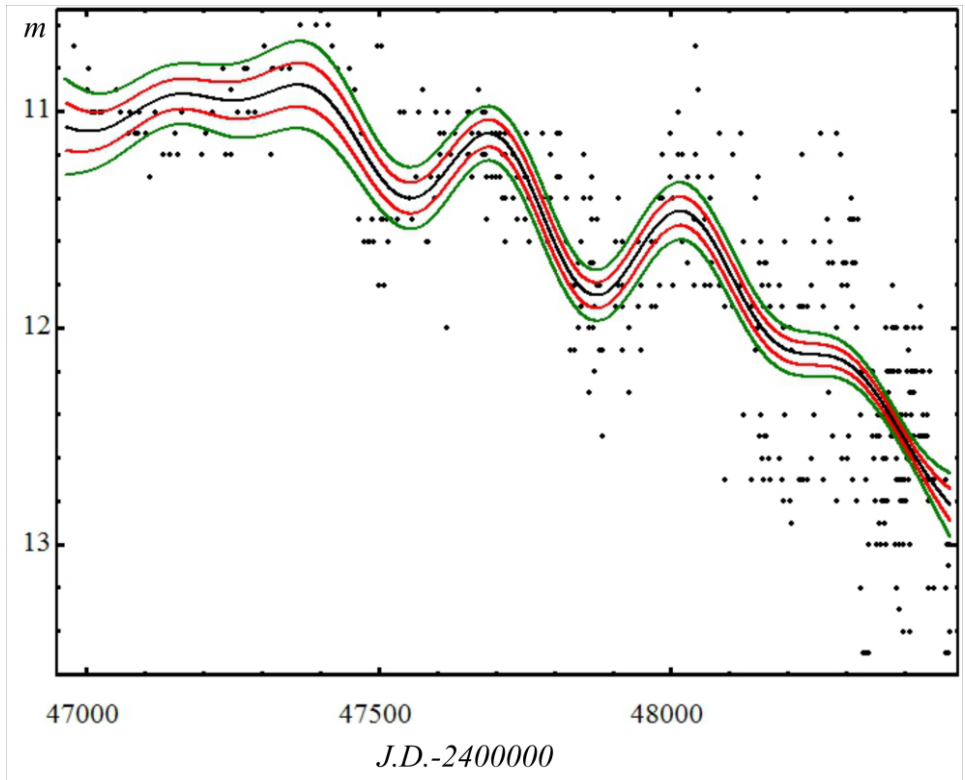

Fig.3. Multifrequency approximation of the first part of the AAVSO's observations of RX Pup by using a sinusoid with periods listed in the Table 2 and the polynomial trend of 2-nd degree.

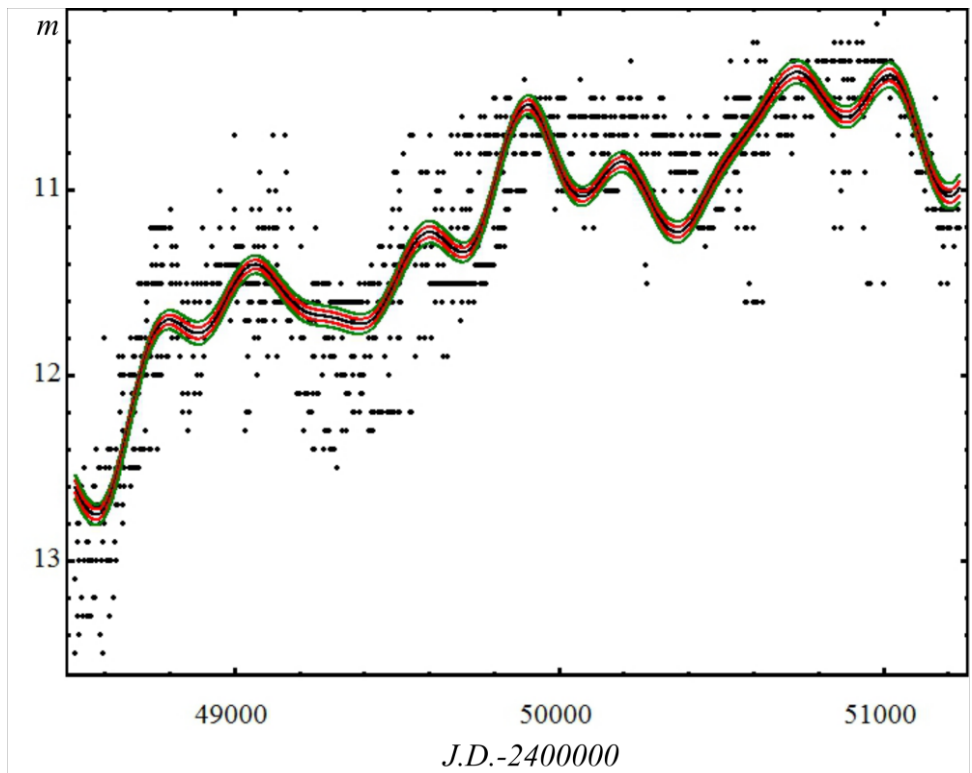

Fig. 4. Multi-frequency approximation of the second part of the AAVSO observations of RX Pup by using sinusoids with periods listed in the Table 2 and the polynomial trend of 2-nd degree.

## AR Pav

AR Pav is the S-type symbiotic variable with regular eclipses of hot component with period of $604.5^d$ (re-determined as $604.75^d$ in our research). The AAVSO observations show (Fig. 5) long-term transition from high to low state (due to accretion activity). In the quiescent phase, Skopal et al. [21] detected the $106^d$-periodic light fluctuations in the visual data and interpret them as pulsations of the red giant.

We rejected the data corresponding to eclipses and made the analysis of non-eclipse light curve (Fig 5). The whole interval give us the period of $631.4^d$ which may be a beat period of $605^d$ and about $14500^d$ of long-term changes in brightness of hot components and its accretion structures. The parameters of approximations with the best periodicities for the quiescent phase (interval J.D. 2454575-2456413) are listed in the Table 3. In all cases, the $604.75^d$-period was used as a primary one with the degree of trigonometrical polynomial $m=5$ to obtain more relevant shape of eclipses and one of the secondary periods listed in the Table 3 with the degree of trigonometrical polynomial $m=1$. Simultaneously, the parabolic polynomial trend was taken into account. In Table 3, $P$ is the period, $S$ is the test function equal to $1-\sigma_{O-C}^2/\sigma_C^2$ [7,10], $S_{O-C}$ is sum of squares of residuals after approximation by using the current period together with $604.75^d$ one and parabolic trend. $S_r$ is equal to semi-amplitude of this oscillation to the its standard deviation and shows the significance of the wave with the secondary period.

The first period may correspond to ellipsoidal variability (due to ellipsoidal shape of the red giant [25]). The $105^d$- periodicity is more significant candidate for two-periodic approximation. But $350^d$-period is also present.

Table 3. Multifrequency approximation of observations of AR Pav at the quiescent phase (interval J.D. 2454575-2456413). See explanation in the text above.

| Periodogram analysis | | Multifrequency approximation | |
|---|---|---|---|
| $P$ (d) | $S$ | $S_{O-C}$ | $S_r$ |
| 605.7±12.7 | 0.174 | – | – |
| 350.7±8.8 | 0.211 | 40.3 | 3.81 |
| 136.3±1.5 | 0.042 | 40.9 | 1.28 |
| 105.1±2.4 | 0.009 | 39.0 | 6.85 |

So we can make the main conclusion that the processes in symbiotic systems require long-term photometrical study and multi-wavelength observations for successful understanding.

This study is a part of the projects "Inter-Longitude Astronomy" [26,27] and "Ukrainian Virtual Observatory" [28].

We sincerely thank variable star amateur observers from AFOEV, VSOLJ and AAVSO for their work that have made such researches possible.

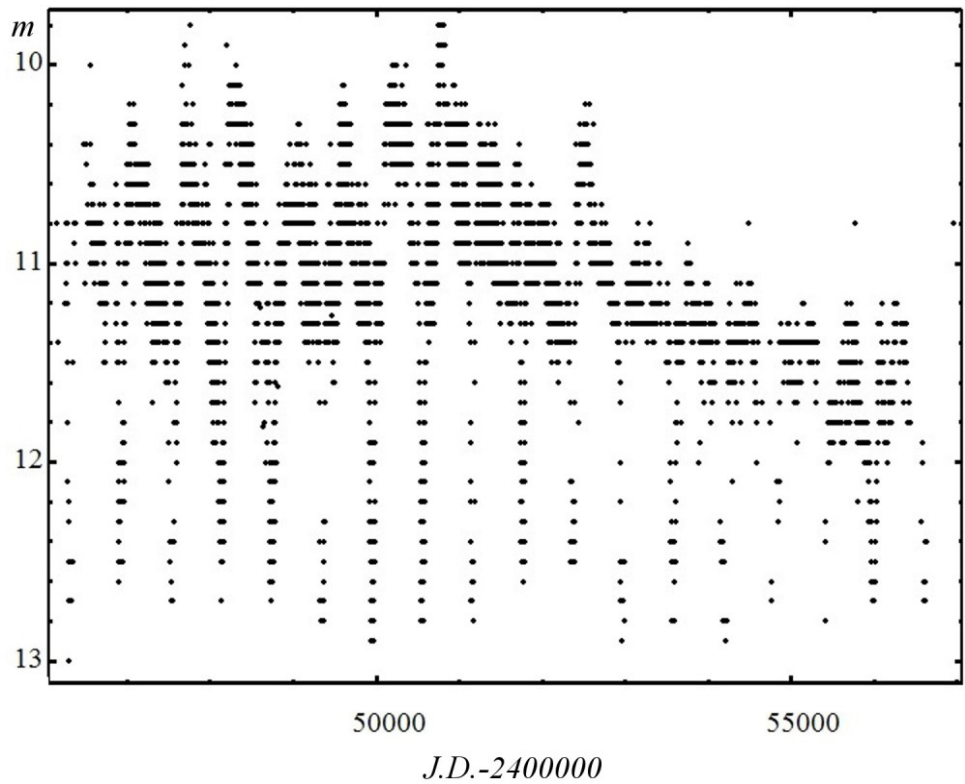
Fig. 5. AAVSO's observations of RX Pup.

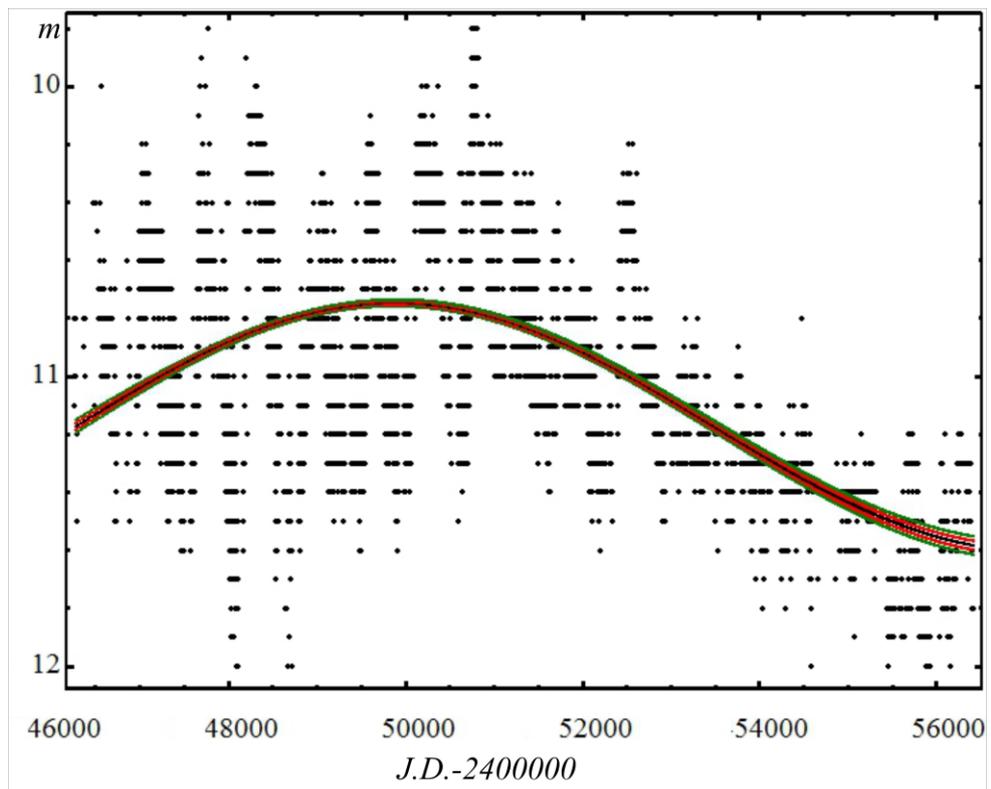
Fig.6. AAVSO's observations of RX Pup without data that cover eclipses and their approximation by using sinusoid with a $14500^d$ period.

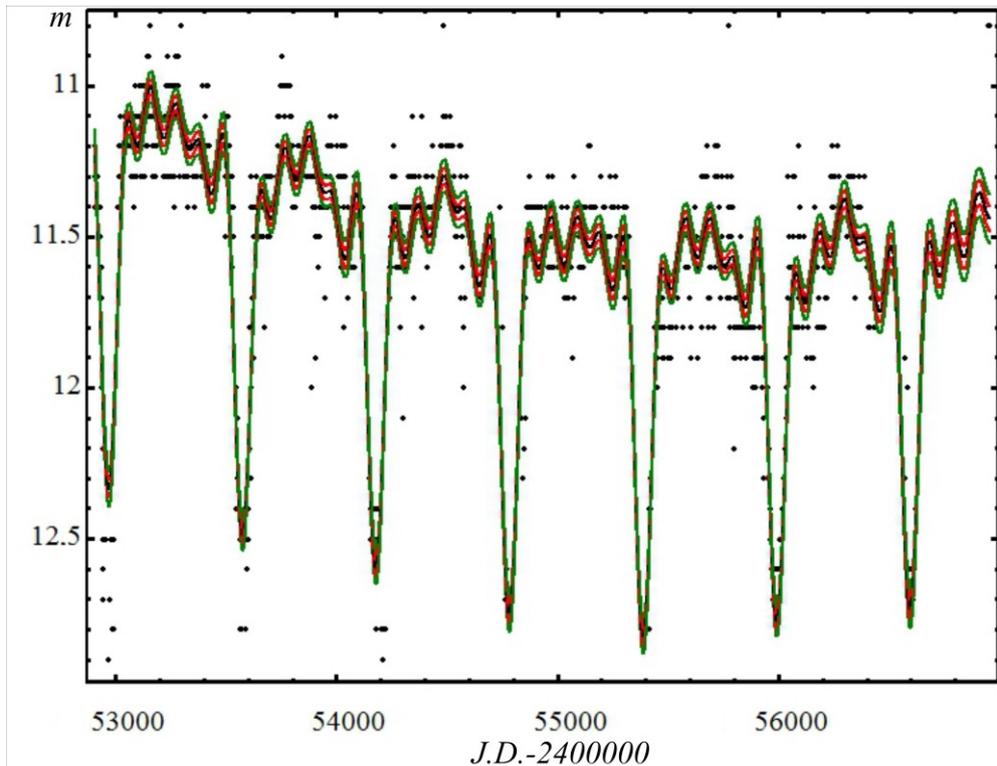

Fig.7. Multifrequency approximation of the part of the AAVSO's observations of AR Pav with $P_1=604.5^d$ (trigonometric polynomial degree $m=5$). $P_2=350.7^d$ ($m=1$) and a superimposed parabolic polynomial trend

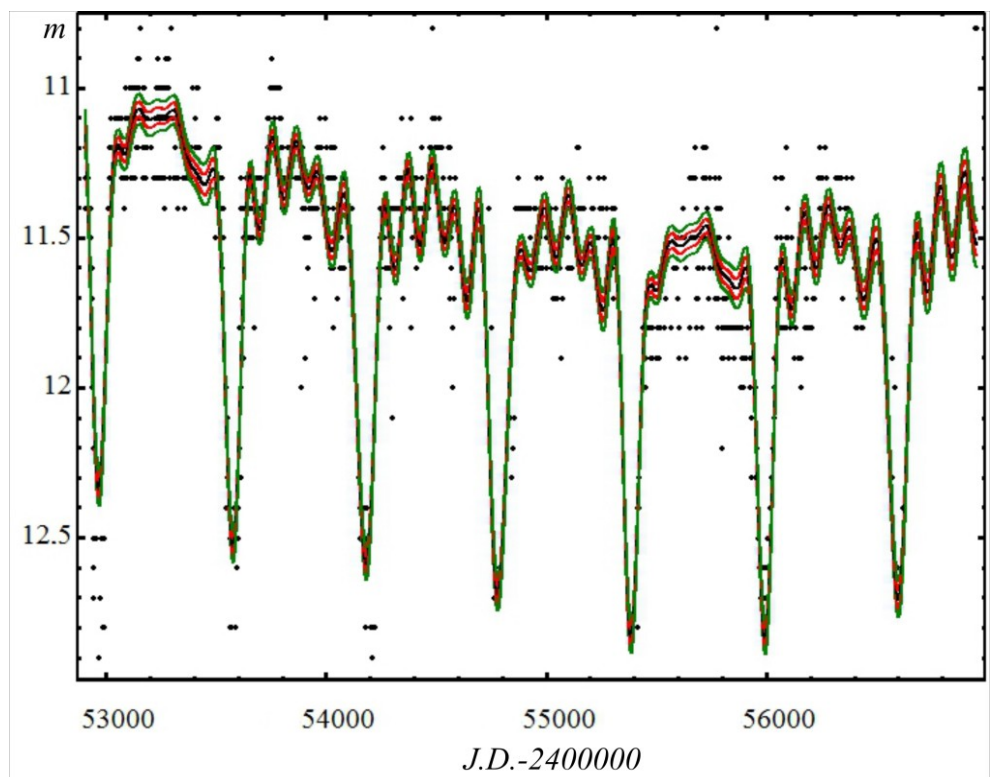

Fig.8. Multifrequency approximation of the part of the AAVSO's observations of AR Pav the $P_1=604.5$ (trigonometric polynomial degree m=5). $P_2=105.1$ (m=1) and a parabolic polynomial trend.

1. Gromadzki M. et al., 2009, Acta Astronomica, 59, 169 (2009AcA....59..169G)
2. Whitelock, P.A. 1987, PASP, 99, 573 (1987PASP...99..573W)
3. Whitelock, P.A. 2003, ASP Conference Series, 203, 41 ( 2003ASPC..303...41W)
4. Hinkle K., et al., 2006, Memorie della Società Astronomica Italiana, 77, 523 (2006MmSAI..77..523H)
5. Mikołajewska J., 2001, ASP Conference Series (IAU Colloquium 183), 246, 167 (2001ASPC..246..167M)
6. Mikołajewski M. et al., 1990, Acta Astr., 40, 129 (1990AcA....40..129M)
7. Andronov I.L., 1994, Odessa Astron. Publ, 7, 49 (1994OAP.....7...49A)
8. Andronov I.L. Marsakova V.I., 2006, Astrophysics, 49. 370 (2006Ap.....49..370A)
9. Kudashkina L.S., Andronov I.L., 1996, Odessa Astron. Publ., 9, 108 (1996OAP.....9..108K)
10. Andronov I.L., 2003, ASP Conf. Ser., 292, 391 (2003ASPC..292..391A)
11. Andronov I.L., Baklanov A.V., 2004, Astronomy School Report , 5, 264
12. Andronov I.L., 1997, Astronomy and Astrophysics Supplement., 125, 207 (1997A&AS..125..207A)
13. Marsakova V.I., Andronov I.L., 1996, Odessa Astron. Publ, 9, 127 (1996OAP.....9..127M)
14. Andronov I.L., Chinarova L.L. 2013, Częstochowski Kalendarz Astronomiczny-2014. P. 171 (2013arXiv1308.1129A)
15. Chinarova L.L., 2010, Odessa Astron. Publ., 23, 25 (2010OAP....23...25C)
16. Andronov I. L., 1998, KFNT, 14, 490 (1998KFNT...14..490A)
17. Mikolajewski M., Mikolajewska J., Khudiakova T. N., 1990, Astronomy and Astrophysics, 235, 219 (1990A&A...235..219M)
18. Andronov I. L., Chinarova L. L. 2003, ASP Conf. Ser., 292, 211 (2003ASPC..292..211A)
19. Corradi L. et al. 2008, Astronomy and Astrophysics, 480, 409 (2008A&A...480..409C)
20. Belczyński I. et al., 2000, Astronomy and Astrophysics Supplement, 146, 407 (2000A&AS..146..407B)
21. Chinarova L. L. 1998, Proc. of the 20th Stellar Conference of the Czech and Slovak Astronomical Institutes, eds.: Dušek J. & Zejda M., Brno, Czech Republic, 37 (1998vsr..conf...37C)
22. Chochol D., Andronov I. L., Arkhipova V.P. et al. 1999, Contributions of the Astronomical Observatory Skalnate Pleso, 29, 31 (1999CoSka..29...31C)
23. Mikołajewska J. et al., 1999, MNRAS 305, 190 (1999MNRAS.305..190M)
24. Skopal A. et al., 2000, MNRAS 311, 225 (2000MNRAS.311..225S)
25. Quiroga C. et al., Astronomy and Astrophysics, 387, 139 (2002A&A...387..139Q)
26. Andronov I.L. et al.: 2010, Odessa Astron. Publ. 23, 8 (2010OAP....23....8A)
27. Andronov I.L. et al.: 2014, Advances in Astronomy and Space Physics, 4, 3-8 (2014AASP....4....3A)
28. Vavilova I.B., et al.: 2012, *Kinem. Phys. Celest. Bodies* **28**, 85 (2012KPCB...28...85V).